\documentclass[11pt]{article}
\usepackage{amsmath}
\usepackage{amsfonts}
\usepackage{graphicx}
\usepackage{hyperref}
\usepackage{url}
\usepackage{xcolor}
\usepackage{xspace}
\usepackage{microtype}
\usepackage{booktabs}
\usepackage{fancyhdr}
\usepackage{lipsum}
\usepackage{siunitx}
\usepackage{cite}
\usepackage{csquotes}
\usepackage{tabularx}
\usepackage{makecell}
\usepackage[utf8]{inputenc} 
\usepackage[T1]{fontenc}
\usepackage[a4paper, left=2.1cm, right=2.1cm]{geometry}
\usepackage[skip=8pt plus1pt, indent=20pt]{parskip}
\title{Migrants as First Responders: A Global Estimate of Disaster-Driven Remittances}

\author{\href{https://orcid.org/0000-0002-3899-1175}{\includegraphics[scale=0.06]{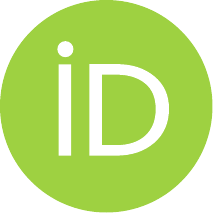}\hspace{1mm}Andrea Vismara} \\
	Complexity Science Hub\\
    Metternichgasse 8\\
	1030 Vienna, Austria \\
	\texttt{vismara@csh.ac.at} \\
    \and
	\href{https://orcid.org/0009-0009-8099-8408}{\includegraphics[scale=0.06]{orcid.pdf}\hspace{1mm}Ola  Ali} \\
	Complexity Science Hub\\
    Metternichgasse 8\\
	1030 Vienna, Austria \\
	\texttt{ola@csh.ac.at} \\
    \and
	\href{https://orcid.org/0009-0000-3859-0825}{\includegraphics[scale=0.06]{orcid.pdf}\hspace{1mm}Carsten Källner} \\
	Complexity Science Hub\\
    Metternichgasse 8\\
	1030 Vienna, Austria \\
	\texttt{kaellner@csh.ac.at} \\
    \and
	\href{https://orcid.org/0009-0009-2261-7726}{\includegraphics[scale=0.06]{orcid.pdf}\hspace{1mm}Guillermo Prieto-Viertel} \\
	Complexity Science Hub\\
    Metternichgasse 8\\
	1030 Vienna, Austria \\
	\texttt{prieto-viertel@csh.ac.at} \\
    \and
	\href{https://orcid.org/0000-0002-0738-2633}{\includegraphics[scale=0.06]{orcid.pdf}\hspace{1mm}Rafael Prieto-Curiel} \\
	Complexity Science Hub\\
    Metternichgasse 8\\
	1030 Vienna, Austria \\
	\texttt{prieto-curiel@csh.ac.at} \\
}

\hypersetup{
pdftitle={Migrants as First Responders: A Global Estimate of Disaster-Driven Remittances},
pdfauthor={Andrea~Vismara, Ola Megahed Ali, Carsten~Källner, Guillermo~Prieto-Viertel, and Rafael~Prieto-Curiel},
pdfkeywords={Human migration, International remittances, Climate-related hazards, Disaster response financing, Adaptation finance},
}

\begin{document}
\maketitle
\begin{abstract}
    
{International remittances represent a vital source of disaster adaptation finance for households around the world, yet their responsiveness to environmental disasters remains poorly quantified. We reveal a previously unmeasured global macro-financial system of international migrant diasporas remittances response to the occurrence of disasters in the country of origin. We do so by developing a structural model simulating individual remittance decisions, calibrated with global disaster records and bilateral monthly remittances flow data from the period 2010-2019. Our analysis reveals that approximately 332 billion USD (5.46\% of total remittances) were mobilized specifically in response to earthquakes, floods, storms, and droughts over the decade. Earthquakes triggered the largest remittance responses per person affected, while droughts elicited the smallest. The model also identifies significant variation in diaspora groups' capacity to activate financial support. These findings establish remittances as a substantial yet limited form of disaster finance, highlighting their importance and limitations in building resilience against future environmental shocks.
}
\end{abstract}

\vspace{1em}

\newpage
\section{Introduction}

{
Every year, billions of dollars flow across borders, sent not from governments or corporations, but from international migrants. These \textit{international remittances} are private financial or in-kind transfers which the current 281 million international migrants send to their families and communities of origin \cite{ratha2017remittances}. Remittances comprise one of the largest global financial flows, reaching an estimated 857 billion USD in 2023, of which 650 billion USD went to low- and middle-income countries \cite{world2024orld}. Remittance flows have been stable and steadily increasing over the last decades and are now higher than the combined total of Foreign Direct Investment and Official Development Assistance \cite{world2024orld}. The United Nations estimates that 800 million people worldwide receive remittances from relatives who migrated abroad \cite{united2019remittances}.  Beyond their volume, remittances matter because they represent one of the most direct links between international migration and development, as they have a positive effect on reducing poverty \cite{adams2005international}, fostering economic growth in countries with less-developed financial systems \cite{sobiech2019remittances}, promoting healthcare access and utilisation \cite{nathaniel2019impact}, and increasing school-enrollment rates \cite{feldmann2025effects}.
}

{
Several theories have been advanced to explain remittance behaviour, with explanations ranging from migrants' desire to provide for their family and community in the country of origin \cite{funkhouser1995remittances, julca2001peruvian}, to self-interest in diversifying income sources for households \cite{stark1988migration,lucas1985motivations} to fulfilling moral obligations to their communities \cite{rapoport2006economics}. Although no unified theory exists, most theories of remittance behaviour imply that remittance flows should respond to shocks in migrants' countries of origin. Indeed, empirical evidence supports this insight by showing that remittance flows are counter-cyclical to income fluctuations in the countries of origin of international migrants, meaning that migrant diasporas abroad mobilise to send more resources home when the economic conditions in the countries of origin deteriorate \cite{frankel2011bilateral}. 
}

{

Remittance flows tend to rise in response to other severe shocks affecting migrants’ families. One of the most significant triggers is the occurrence of natural hazards that escalate into disasters in migrants’ countries of origin. Examining the links between natural hazard-related disasters and remittances is increasingly relevant. Climate change is driving a measurable rise in the frequency and intensity of hazards such as storms, floods, and droughts, which has increased the frequency of disasters, defined as the social and economic consequences arising when hazards interact with exposure and vulnerability \cite{thomas2015global}. On the economic side, the last decades have seen an increase in losses and damages from disasters, especially those associated with extreme events \cite{coronese2019evidence}. In 2024 alone, the impacts of natural hazard-related disasters generated an estimated 328 billion USD in damages, well above the ten-year average of 254 billion USD, and caused over 14,000 fatalities worldwide \cite{sigma_report_25}. Global losses from disasters have grown not only because hazard events are intensifying but also because more people and high-value assets are now located in risk-prone areas, thereby increasing exposure and risk \cite{guneralp2015changing}. In this context, disaster impacts can create acute shocks at the household level, making remittances an important mechanism for immediate adaptation and post-event recovery, especially in countries where disaster adaptation infrastructure remains underdeveloped.
}

{
Remittances contribute positively to building resilience in migrants' communities of origin both before and after disasters occur \cite{sikder2017remittances}. In terms of preparedness, remittances provide funds for infrastructure investments, serving as an alternative source of adaptation finance \cite{manic2017impact}. Remittances can thus close adaptation gaps, especially reaching the most vulnerable to disaster impacts, whether these are related to climate change, such as droughts  \cite{musah2018migrants}, or not, such as earthquakes \cite{manandhar2016remittance}. With regard to post-disaster recovery, access to remittances can improve the well-being of families and communities affected. On an aggregate level, remittance flows are associated with a reduction in poverty in post-disaster periods \cite{mbaye2017natural}. Families who can rely on remittances show higher consumption patterns in post-disaster scenarios, as well as being protected from having to sell household assets or livestock to cope with the disaster's effects \cite{mohapatra2012remittances}. Receiving remittances has also been associated with other positive effects, such as alleviating the negative impacts of disasters on mental health \cite{tachibana2019remittances}. 

In the broader context of climate change risks and migration, the existence of remittance flows is considered one of the criteria that determine the success of migration as an adaptation response \cite{szaboova2023evaluating}. The international and academic communities are also increasingly concerned with understanding how to mobilise private financial flows in complementing public resources for disaster response and climate adaptation \cite{phillips2023disaster}. Remittances potentially play a critical role at the intersection of these discussions. A better understanding of remittances flows in the aftermath of disaster events would bring an important contribution to the development of climate change and natural hazard adaptation. However, despite their prominence, there is very limited empirical evidence on the extent to which remittance flows are mobilised specifically in response to disasters, as opposed to reflecting regular or long-term transfer patterns.

}

{
Remittances tend to increase in the aftermath of natural disasters \cite{david2011international, mohapatra2012remittances}. Evidence shows that remittance senders tend to rush to provide assistance after a disaster, but since their financial situation does not change, the immediate increase is compensated later, resulting in a short-term increase following the onset of disasters, but small yearly effects \cite{bragg2018remittances}. A recent study based on monthly remittance flows from Italy shows an immediate positive response of remittances to disasters in migrants’ home countries, which persists for up to 4 months after the disaster \cite{bettin2025responding}. However, other research has found no significant effect of disasters on remittance increases \cite{yang2008coping, lueth2008determinants}.

Discrepancies in the empirical evidence on disasters and remittances largely stem from limitations in both data and methodology. Most quantitative studies rely on econometric analyses combining quarterly or yearly panel data on remittance flows with disaster indicators \cite{zhou2023review}. Such temporal aggregation is poorly suited to capturing disaster-driven responses, which are inherently time-critical. In this sense, the availability of high frequency temporal data is crucial for detecting effects. This limitation is compounded by data constraints: there is no comprehensive dataset of bilateral remittance flows at high temporal resolution. The most widely used alternative, the World Bank’s KNOMAD bilateral remittances matrix, provides model-based annual estimates derived primarily from migrant stocks and GDP differentials \cite{KNOMAD_Remittances, ratha2007south} (see Supplementary Material, section \ref{sec:KNOMAD}). By construction, this approach cannot capture short-term behavioural responses to disasters and may introduce systematic biases through the assumption that all migrants remit. On the methodological side, regression frameworks further restrict inference by imposing linear, equilibrium-oriented relationships that are ill-suited to the nonlinear and adaptive decision-making processes that characterise remittance behaviour \cite{klein2018agent, crutzen2023regression}.
}

{
For these reasons, we do not yet have a clear picture of how much money migrant diasporas mobilised through remittances in response to disasters. Our work aims to address this gap. To do so, we contribute a novel structural model, where individual remittance decisions are conditional on migrant characteristics and the occurrence of disasters in the countries of origin of migrants. The model is unique because it is built upon the assumption that individual migrants make a binary decision to remit at every period, conditional on micro-level factors and the macro-level shock of a disaster. The model is trained using monthly bilateral remittance data and exploits the exogenous variation in disaster timing and severity to quantify the remittance response. We include four types of disasters in our analysis, namely floods, droughts, storms, and earthquakes, based on the information recorded in the Emergency Events Database (EMDAT, \cite{delforge2025dat}). The structural model is used to estimate a global matrix of  bilateral remittance flows for the 2010-2019 period.   
}

{
Our approach, which embeds age-specific earnings-to-consumption patterns, family ties, income and GDP differentials, and disaster magnitude, accurately reproduces historical bilateral remittance dynamics, including both the surge and the gradual decay of post-disaster remittances. We find that between 2010 and 2019, international migrants transferred a total of 6.1 trillion USD. Of this sum, 5.46 percent was mobilised in response to disasters, equivalent to 332 billion USD. Around 90\% of the total-disaster induced remittances went to lower- and upper-middle-income countries. Floods are the disaster category which mobilised the largest total amount of remittances, with a total of 138 billion USD, while earthquakes are the disaster which mobilised the largest sum of remittances per person affected. 
}


\begin{figure}[hbt!]
    \centering
    \includegraphics[width=\textwidth]{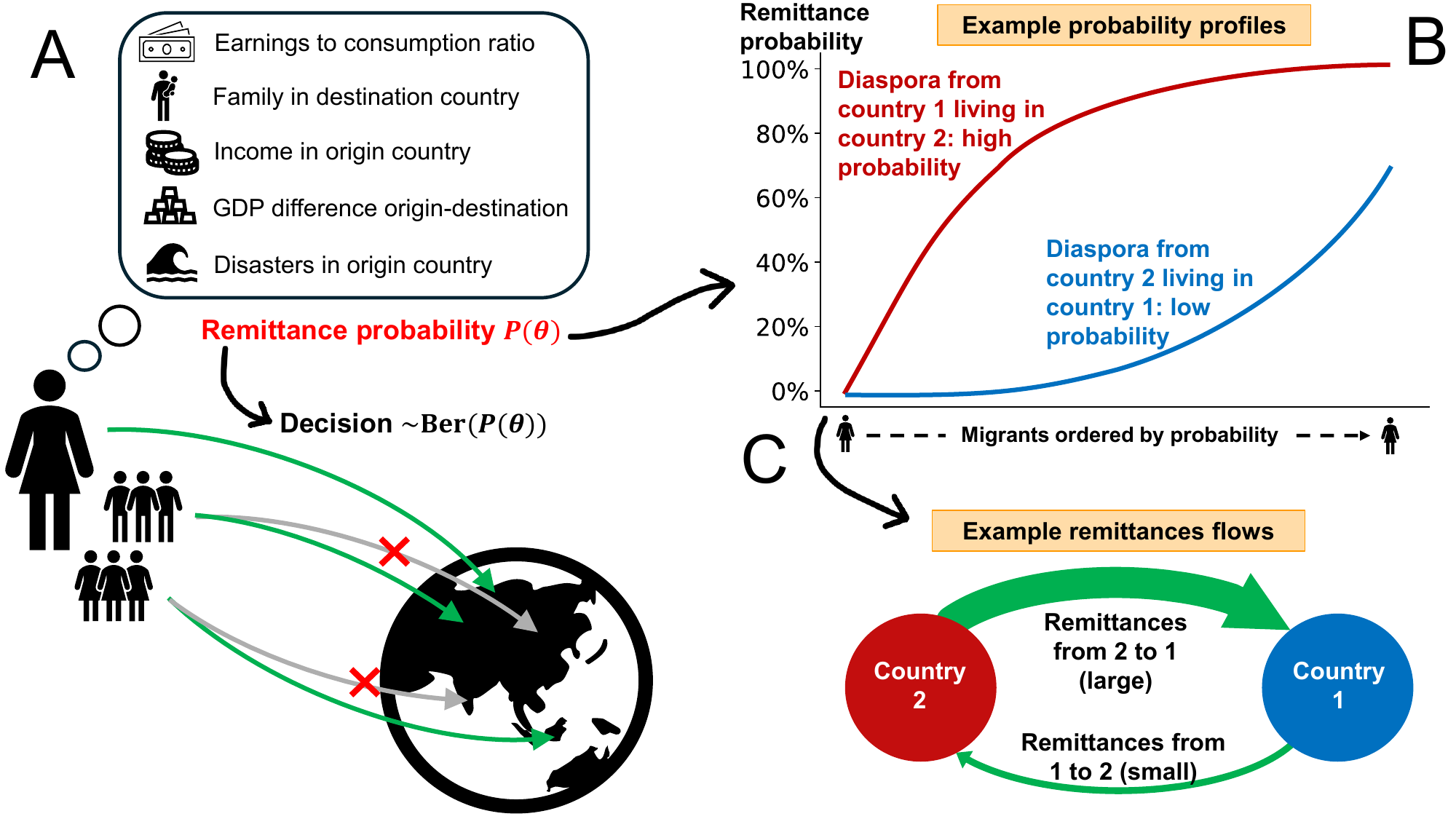}
    \caption{\textbf{Model description and sample results.} Panel A: schematic representation of the structural model, listing the factors that influence a migrant's remittance probability. The probability determines how many migrants send remittances to their countries of origin. Panel B and C: sample probability profiles, which determine the bilateral remittance flows.}
    \label{fig:model_introduction}
\end{figure}

\section{Results}

\subsection{A structural model of remittance-sending decisions}
{
{
We build a structural model that captures international migrants' decisions about sending remittances. Explicitly modelling the individual remittance choices is to the best of our knowledge a novelty in the literature. Our model assumes that each migrant makes a binary Bernoulli choice at every period which can result in a \enquote{success}, meaning sending remittances in this period or a \enquote{failure}, meaning no remittances sent. The probability for this choice depends on i) the migrant's age-dependent ratio of earnings to consumption, which depends on their age; ii) the probability of having family in the country of destination; iii) the difference in GDP between origin and destination country; iv) the income level of the country of origin; v) the occurrence and magnitude of disasters in the origin country (Figure \ref{fig:model_introduction}, A). From individual remittance choices, we derive aggregate country-to-country remittance flows over time. We calibrate the model to match a panel of monthly, country-to-country remittance flows for the 2010-2019 period. This reference dataset covers a total of 769 billion USD in bilateral remittance flows, equivalent to around 12.5\% of the total remittances moved over the period, and it contains information on 194 unique sending countries and 140 unique receiving countries (more details in the Methods, section \ref{sec:rem_data}). 

}

{
In our model, all international migrants have a probability of sending remittances, which changes over time. We aggregate migrants into national diasporas. A diaspora comprises all the people who live outside the country in which they were born. Collecting and ordering the remittance probabilities of all individuals in a given diaspora allows us to produce \textit{diaspora remittance probability profiles}, which summarise each diaspora's remittance-sending behaviour. This allows us to distinguish between diasporas with high and low remittance-sending probabilities, as well as the distribution of these probabilities within the population (Figure \ref{fig:model_introduction}, B). From the probability profiles, we extrapolate the remittance flows between countries (Figure \ref{fig:model_introduction}, C). To derive the flows, we assume that migrants who decide to send money remit a fixed portion of their monthly income. The parameter for this fixed portion is also calibrated. 
}

{
With the calibrated model, we derive estimates for the probability profiles of global diasporas (Figure \ref{fig:profiles}, A). The shape of the probability profiles also explains the potential for a diaspora to send more remittances during a disaster. We call this potentiality \textit{activation capacity}. What the model captures is that diasporas where most people are already sending remittances will not activate in response to a shock (e.g. the diaspora from the Philippines), as well as with diasporas with a low probability of sending remittances: the occurrence of a disaster will not push many towards activating (e.g. the diaspora from Finland). Only diasporas where most individuals have a median probability, such as the diaspora from El Salvador, will have high potential for activation in response to disasters. 
}

{
We also analyse the temporal trends in bilateral remittance flows by tracking the evolution of the remittance probability profiles for national diasporas living in specific countries (Figure \ref{fig:profiles}, B). The model captures effects beyond a simple linear relationship between the number of migrants and total remittance flows. A relevant example of this dynamic is the remittances flow from the USA to Mexico. Over the past decade, the number of Mexican migrants in the USA has declined \cite{mazza2017us}, yet remittances have continued to grow in absolute and real terms \cite{cardenas2025}. At first glance, this seems paradoxical, since one would expect fewer migrants to mean less money sent home. Our model, however, reproduces the growing remittances trend despite the decline in diaspora size and suggests that the apparent contradiction is explained by a higher probability of sending remittances. Two factors contribute to this rise: the strengthening of the USA economy relative to Mexico and shifts in the diaspora’s composition, with a growing share of single individuals who, in our framework, tend to send remittances more frequently.  
}

\begin{figure}[hbt!]
    \centering
    \includegraphics[width=\textwidth]{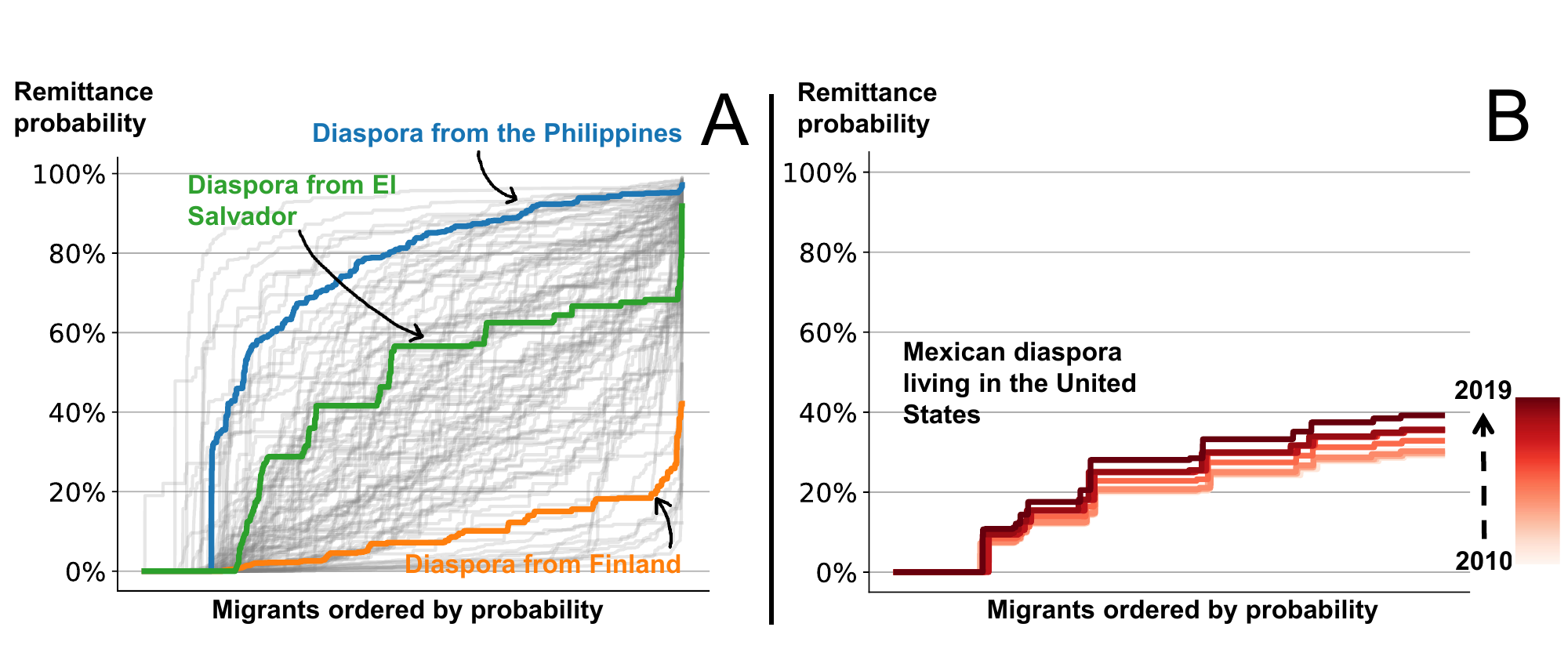}
    \caption{\textbf{Probability profiles of sending remittances.} Panel A: probability profiles for international diasporas of more than 500,000 people in 2019, by country of origin, irrespective of destination country. Panel B: evolution of the remittance probability profile of the Mexican diaspora in the United States (2010 to 2019).}
    \label{fig:profiles}
\end{figure}

}

\subsection{Demographic composition of remittance senders}

{
We estimate that around 56\% of remittance senders between 2010 and 2019 are male. This number is higher than the average proportion of male international migrants over the decade, which is 52.7\%. There are important differences between the countries of origin of migrants. For migrants from low and lower-middle income countries, the percentage of male senders is closer to 60\%, while for high and upper-middle income countries it is less than 50\%. These numbers largely reflect the demographic composition of international migrant diasporas (see Supplementary Material, section \ref{sec:migrants}).

The average age of remittance senders globally is around 37. There is significant variation across countries of origin of migrants: the average age of a remittance sender who migrated from a low-income country is 33 years old, while the average age of a sender coming from a high-income country is 40 years old. In general, 60\% of remittance senders worldwide are between 20 and 39 years old. Among migrants coming from low-income countries, up to 75\% of remittance senders are below 40 years of age. On the contrary, for migrants coming from high-income countries, only 52\% of remittance senders are below 40. The age distribution of senders also largely reflect the demographic distribution of migrants globally.
}
\vspace{3em}

\subsection{Global structure and distribution of remittance flows}
{
{
With our structural model, we estimate a global matrix of country-to-country monthly remittance flows (Figure \ref{fig:remittances_matrix}, A). Within these flows, we can distinguish between the size of regular and disaster-induced remittances. Our estimates show that, between 2010 and 2019, international remittance flows across the globe totaled 6.1 trillion USD. This sum is equivalent to 0.78\% of the global GDP produced over the same time period and is in line with the aggregate remittances estimated by the World Bank \cite{world2024orld}. High-income countries are the largest senders overall, and flows tend to be clustered among the same-income-group countries, especially for remittance senders living in lower-middle- and low-income countries.
}

\begin{figure}[hbt!]
    \centering
    \includegraphics[width=\textwidth]{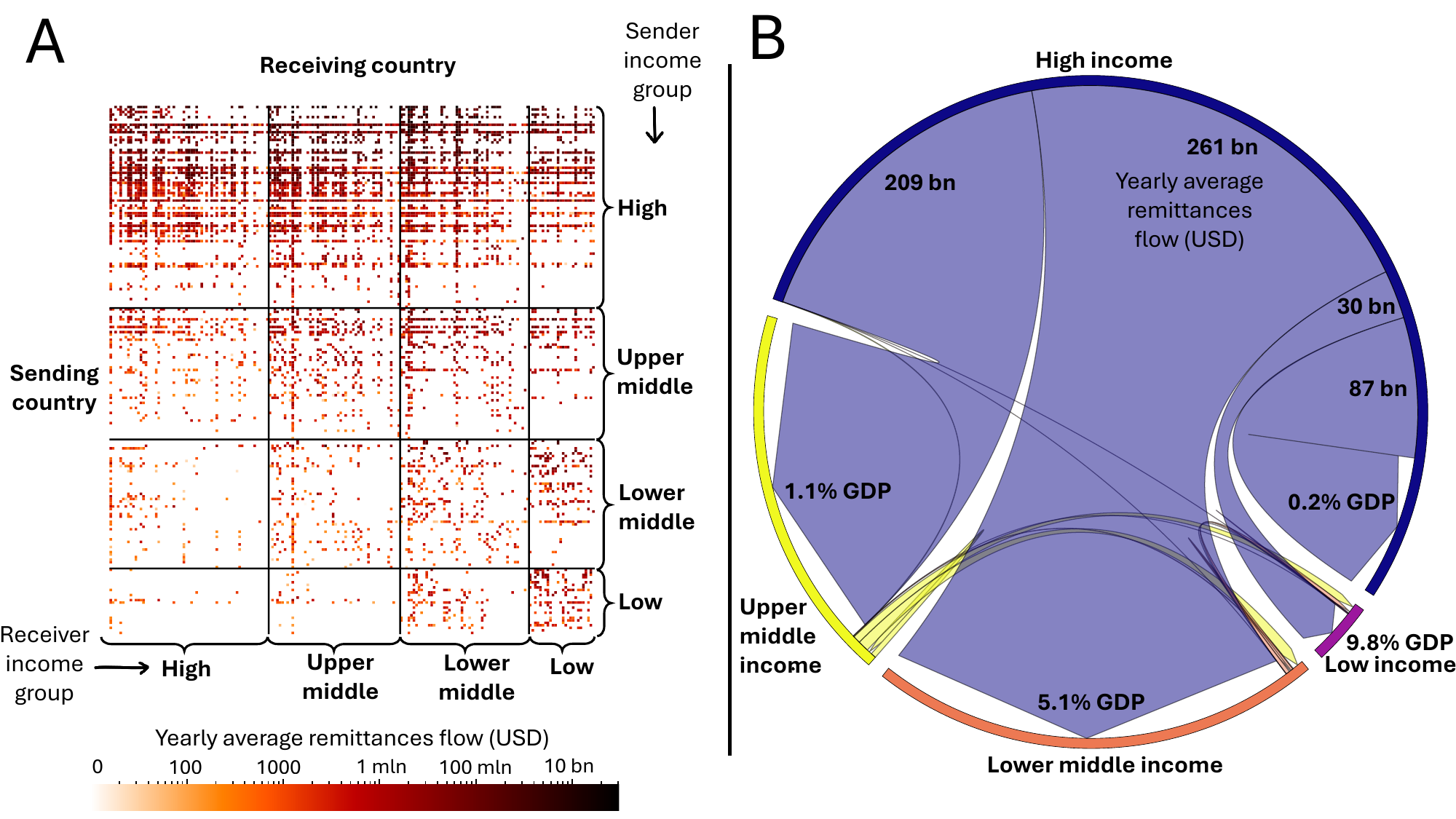}
    \caption{\textbf{Estimated bilateral remittance flows.} Panel A: matrix of bilateral remittance flows on an average year (in US dollars). Panel B: yearly average flows between blocks of countries as well as the total volume in USD and the percentage of GDP contributed on the receiving side for remittance flows from high-income countries.}
    \label{fig:remittances_matrix}
\end{figure}

{
Around 96.5\% of the total flows are sent by international migrants living in high-income countries (Figure \ref{fig:remittances_matrix}, B). To put it in context, only 72\% of international migrants live in high-income countries.  Migrants in upper-middle-income countries contribute another 2.5\% of the total remittance flows, while the rest is sent from lower-middle and low-income countries. On the receiving side, the share is split more evenly, with high-income countries receiving 14\% of total remittance flows, while 35\% and 44\% go to upper-middle- and lower-middle-income countries, respectively. Only 6\% of international remittance flows go to low-income countries.
}
\vspace{2em}

{
These numbers, however, hide the contribution that remittances have relative to the wealth of countries. In total, 1.1\% of the GDP of all high-income countries was transferred through international remittances between 2010 and 2019. The same sum of money contributed 0.2\% of high-income countries' GDP, 1.1\% of upper-middle income countries' GDP, 5.1\% of lower-middle income countries' GDP, and 9.8\% of low-income countries' GDP combined (Figure \ref{fig:remittances_matrix}, B). 
}

{
Latin America, Eastern Europe, North Africa, and Central Asia received the highest remittance per capita (Figure \ref{fig:disaster_remittances_map}, A). According to our estimate, El Salvador, Guyana, and Suriname are the three countries that receive the highest amount of remittances per year, with more than a thousand dollars per inhabitant on average. Almost all countries in Eastern Europe also receive large remittance flows per capita due to high labour migration to richer European countries. Countries in central and Sub-Saharan Africa receive relatively small remittance inflows because most migration from the region occurs within short distances, towards countries with low income levels. 
}

\begin{figure}[hp]
    \centering
    \includegraphics[width=0.9\textwidth,height=0.95\textheight,keepaspectratio]{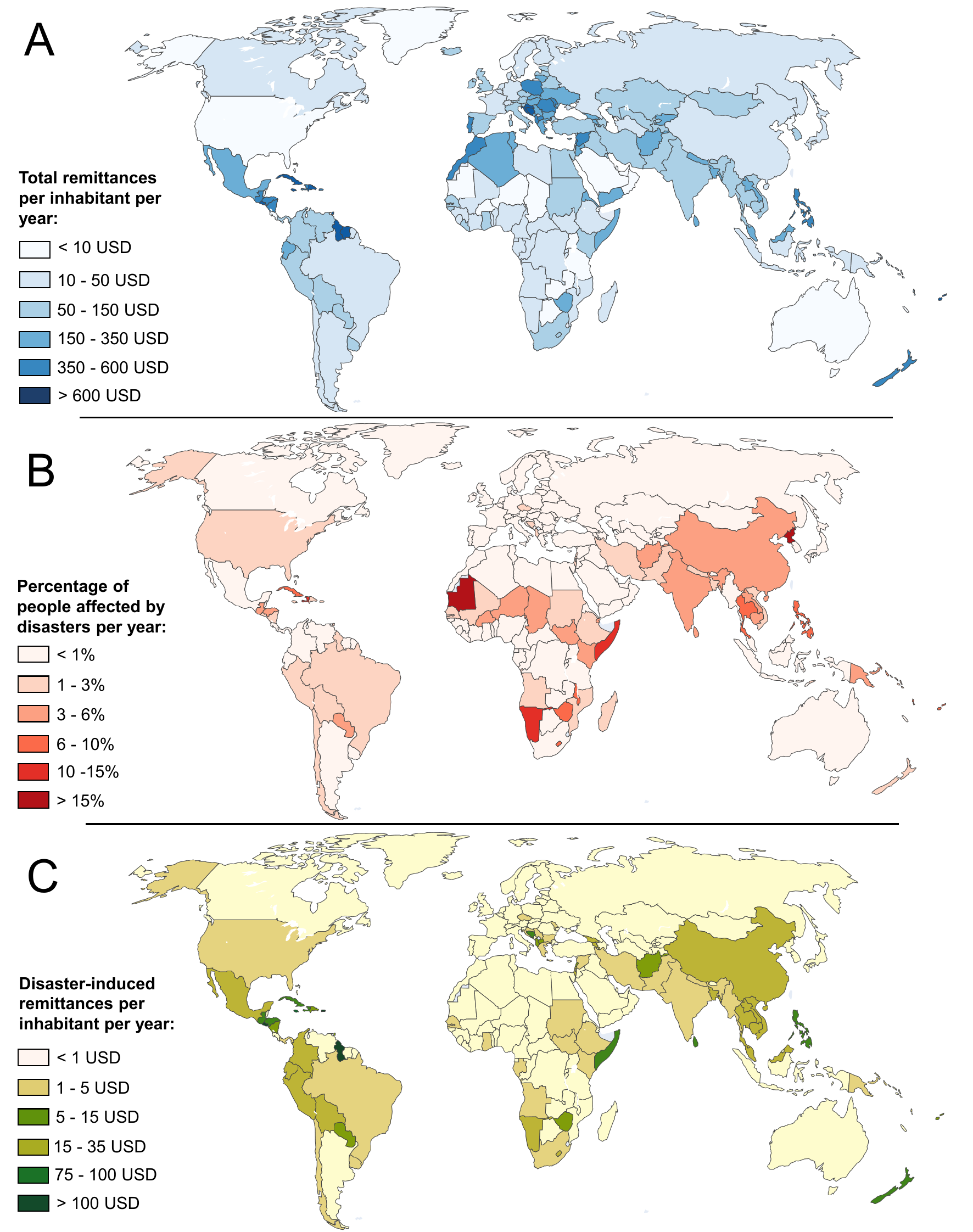}
    \caption{\textbf{Total remittances, disasters and disaster-induced remittances (2010-2019).} Panel A: yearly average remittances per inhabitant received by country. Panel B: yearly average share of the country's population that was affected by disasters (floods, storms, droughts, earthquakes). Panel C: average yearly disaster-induced remittances per inhabitant received by country.}
    \label{fig:disaster_remittances_map}
\end{figure}

}

\subsection{Magnitude and dynamics of disaster-induced remittances}
{
{
We obtain information about disasters' occurrence and impact from the Emergency Events Database (EMDAT). For the scope of our analysis we concentrate on floods, storms, earthquakes, and droughts. EMDAT collects information on any event that crosses any of three thresholds: ten fatalities, 100 people affected, or the declaration of a state of emergency \cite{delforge2025dat}. We exclude heatwaves despite their extensive mortality due to difficulties in comparing the effect to other disasters (see Supplementary Material, section \ref{sec:disasters}). According to EMDAT, between 2010 and 2019, there were around 3,000 disaster events connected to the occurrence of floods, storms, earthquakes, and droughts. These events affected a total of 1.74 billion people, the vast majority of whom lived in lower-middle-income and upper-middle-income countries (Figure \ref{fig:disaster_remittances_map}, B). 
}

{
We estimate that the occurrence of a disaster, on average, increases remittance sending for up to nine months after the event, with a peak around the fourth month after the event, and a small negative effect after the ninth month. The disaster-induced flow of international remittances between 2010 and 2019 amounted to circa 332 billion USD, equivalent to 5.46\% of total remittance flows. These estimates are conditional on the reported disaster effects in EMDAT. Mirroring the distribution of disaster events, the largest recipients of disaster-induced remittances were lower- and upper-middle-income countries, with roughly equal shares of 44\% of the total disaster-induced remittances over the 2010 to 2019 period. The three years with the most significant movement of disaster-induced remittances were 2016, 2011, and 2017, with a movement of 47, 41, and 39 billion disaster-induced remittances, respectively.
}

{
The five countries that received the largest absolute amounts of disaster-induced remittances are, in order, China, India, the Philippines, Mexico, and Bangladesh. These countries combined received around 186 billion USD in disaster-induced remittances over the 2010 to 2019 period, which is equivalent to 56\% of the total. The relative amount of disaster-induced remittances received by each country largely reflects the number and magnitude of disaster events that hit the country, while the total amounts also depend on the characteristics of the diasporas abroad. All of these countries are estimated to have more than 6 million migrants living abroad. 

\vspace{2em}
However, there are notable variations in the probability of sending remittances of each diaspora, and consequently in their capacity to activate in response to disasters. For example, the United States is the fourth country by number of people affected by disasters, while ranking only nineteenth in disaster-induced remittances (4.2 billion USD). On the contrary, countries like Mexico and El Salvador suffered less from disasters but could mobilise large amounts of remittances per inhabitant (Figure \ref{fig:disaster_remittances_map}, C). Another notable dynamic is that for some countries, the portion of remittances induced by disasters is especially large. In Somalia, Zimbabwe, and Namibia, more than 20\% of remittance inflows are linked to disaster response.    
}

\subsection{Notable disaster events}
{
Our modelling approach also allows us to separate the portion of remittance flows connected to the occurrence of a single disaster event. The three most significant single-disaster episodes by the amount of remittances mobilised were all flood-type disasters that occurred in China in 2010, 2011, and 2016. All three events occurred between May and June, lasted for months, and affected more than 60 million people each, according to EMDAT. The three disasters induced an additional flow of remittances of 16, 14, and 11 billion USD, respectively over the twelve months following the disaster.

We now turn to three of the most well-known disasters which took place over the 2010-2019 decades. The 2010 Haiti earthquake was a devastating disaster with a death toll of more than 160 thousand people \cite{kolbe2010mortality}, making it the deadliest disaster of the 21st century for a single country. According to our estimates, remittances to Haiti increased by around 20\% over the year following the earthquake, generating a bump in remittances of 713 million USD over the normal flow. This estimate is almost double the 320 million USD extra predicted by the World Bank \cite{ratha2010outlook}. Typhoon Haiyan is a catastrophic tropical cyclone which struck South-East Asia in November 2013, affecting especially the Philippines, where it caused more than 6 thousand deaths \cite{lagmay2015devastating}. According to a survey based study, around 4\% of families with relatives abroad started receiving remittances after the disaster \cite{cabuay2025insuring}. Our model estimates a 15\% increase in remittance flows over the year following the typhoon, equivalent to 6.6 billion USD in extra remittances flowing to the Philippines. Lastly, the April 2015 earthquake in Nepal killed around 9000 people. The World Bank estimates a 20\% year-on-year increase in remittances in the financial year following the earthquake \cite{ratha2016migration}, but there is no estimate of the amount of disaster-induced remittances. Moreover, the World Bank estimate number does not differentiate between money sent by migrants and other forms of international solidarity. According to our estimate, an extra 490 million USD in migrants' remittances flowed to Nepal over the 12 months following the earthquake, equivalent to a 7\% increase over the baseline flow. 
}
}

\subsection{Remittance mobilization across disaster types}
{
{
Floods moved the largest amount of remittances, with a total of 138 billion USD (Figure \ref{fig:disaster_flows}, A). The large mobilization is due to the combination of size and frequency of flooding events with their occurrence in countries with large international diasporas, such as China, Pakistan and Bangladesh. Droughts induced the second-largest flow of remittances, totaling 96 billion USD. Droughts occurred at a less consistent rate than floods, as reflected in the time series of drought-induced remittances. The widespread period of droughts between 2015 and 2017 in Latin America, Sub-Saharan Africa, and Eurasia is reflected in remittance flows. Earthquake-induced remittances were also unstable: a large spike at the beginning of 2018 was driven by the earthquakes that affected Mexico in 2017 and Indonesia in 2018.
}

{
Not all disasters generate the same remittances mobilisation (Figure \ref{fig:disaster_flows}, B). Comparing total disaster-induced remittances with the number of people affected for each disaster category shows that earthquakes accounted for the largest relative amount, with 542 USD per affected person. Earthquakes are sudden and cause large impacts, and have occurred in countries with diasporas that could be activated. On the contrary, droughts caused the smallest relative impact, with 142 USD per affected person. Droughts are a \textit{creeping phenomenon}: their effects accumulate slowly, and they last for prolonged periods of time \cite{wilhite2016drought}. For this reason, migrant diasporas cannot sustain sending higher amounts of remittances for events that last long periods. The differences in remittances induced by disasters are also explained by the countries affected and the location of their migrant diasporas, as diasporas in richer countries can send larger sums.
}

\begin{figure}[hbt!]
    \centering
    \includegraphics[width=\textwidth]{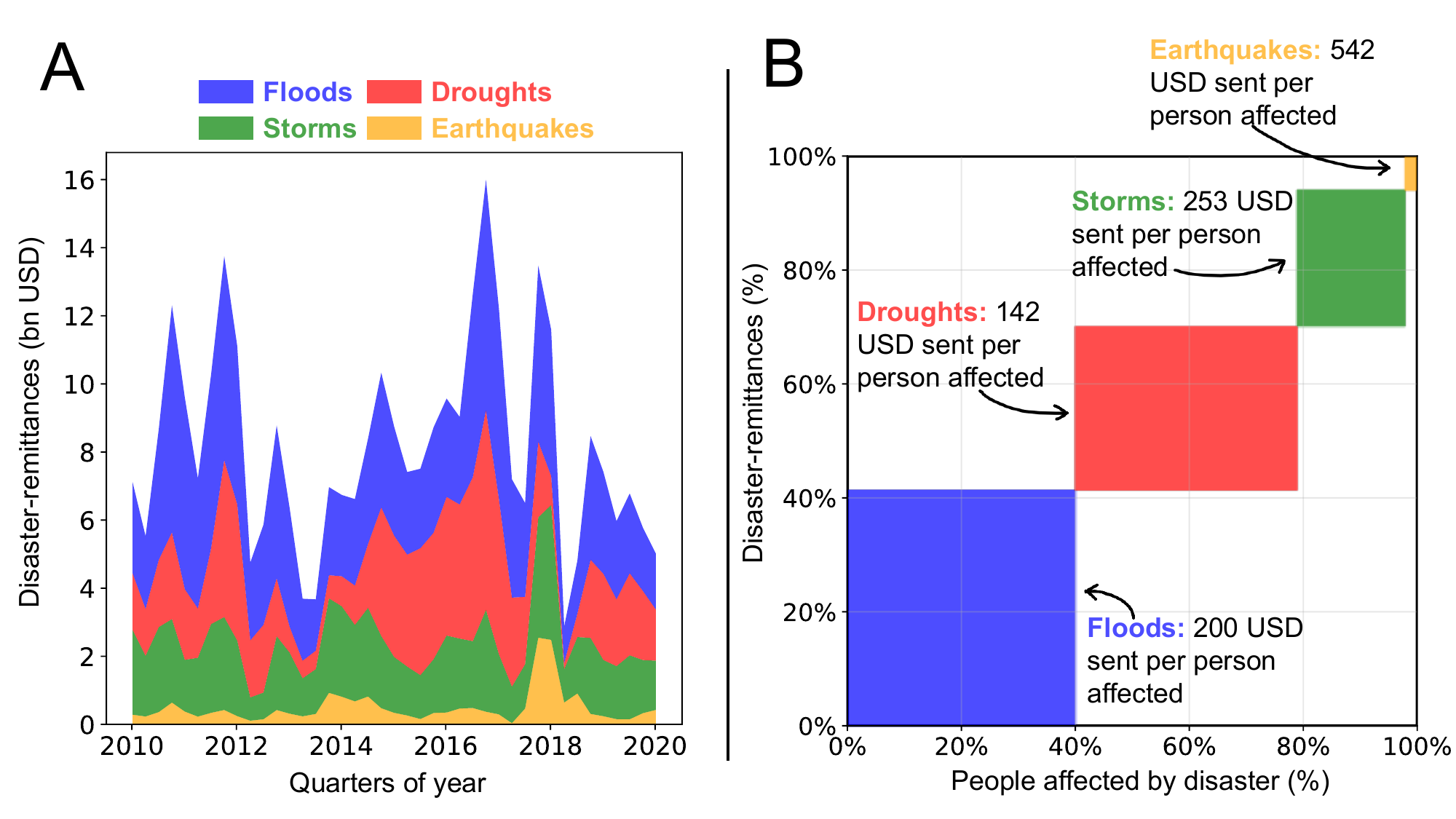}
    \caption{\textbf{Disaster remittances by disaster type}. Panel A: volume of remittances induced by each disaster category, grouped by quarter of the year. Panel B: relationship between the number of people affected (x-axis) and the amount of remittances mobilised (y-axis) by disaster category.}
    \label{fig:disaster_flows}
\end{figure}
}

\section{Discussion}
{

{
Our analysis provides new evidence on the role of remittances as a financial response mechanism to sudden-onset disasters. By developing and calibrating a structural model of individual remittance behavior, we provide new estimates of monthly bilateral remittance flows between countries for the 2010–2019 period, including remittances in response to disasters. We estimate that approximately 5.46\% of global remittances over the 2010–2019 period were mobilised in direct response to natural disasters. This amount, equivalent to circa 332 billion USD, highlights that migrant diasporas collectively constitute a significant, though uneven, source of disaster finance. 
}

{
A first insight we highlight concerns the heterogeneities of disaster-induced remittances. Our estimates show that middle-income countries are the largest recipients of disaster-induced remittances, while low-income countries benefit relatively little in absolute terms. The amount of remittances moved in response to disasters is not directly correlated with the number and magnitude of disasters suffered by the country of origin of migrants, but is mediated by the characteristics of migrant diasporas abroad. For example, the USA suffered a high number of disasters over the decade, but received relatively little money in disaster-induced remittances, due to the low activation of migrant diasporas from the US. On the contrary, countries like China, India, and Pakistan have international migrant diasporas who show high activation capacity in response to disasters. 

In general, migrants from low income countries tend to have the highest probability of sending remittance \textit{before} disasters occur. This limits their capacity to activate more financial resources after the occurrence of a disaster. Similarly, migrants from high income countries have lower baseline probabilities of sending remittances, and therefore lower capacity to activate after disasters. Analyzing these patterns is important because it can uncover hidden vulnerabilities. For example, the households most vulnerable to disasters are often located in low-income countries. Understanding diaspora conditions is key to understanding the potential for remittances to generate disaster finance, without assuming that all international migrants will be able to mobilise in the same way.

A second insight we highlight is the heterogeneity in disaster-induced remittances across hazard types: earthquakes generated the largest relative inflows per affected person, while droughts generated the smallest. This result can be understood as reflecting both the suddenness and high magnitude of earthquakes in terms of the number of people affected, and the usually prolonged nature of droughts, which make sustained higher transfers difficult for migrant households to maintain. Droughts also disproportionately affect low-income countries, creating an additional challenge for disaster adaptation. Floods emerged as the most significant absolute driver of remittances, mainly because they are frequent and heavily affect countries with large overseas diasporas. These differences underscore that remittances are not a uniform coping mechanism but vary depending on hazard characteristics, exposure, and diaspora location and condition. 

}

{
Our results are also consistent with previous evidence showing a temporary surge and gradual dissipation of remittance inflows following disasters \cite{bragg2018remittances, bettin2025responding}. The estimated nine-month duration before the effects taper off suggests that while remittances can support short- and medium-term disaster recovery, they cannot fully substitute the sustained capital investments required to build adaptive capacity and facilitate transformational adaptation for long-term climate resilience \cite{tong2021characteristics}. This creates a double constraint for low-income countries: given that disaster-induced remittances surges are both temporary and harder for low-income countries to access, they cannot replace other instruments of adaptation finance. This reinforces the need for targeted public mechanisms, such as Loss and Damage funds, to fill the structural resilience gap that private financial flows leave behind \cite{serdeczny2023research}.
}

{
Our structural model relies on monthly bilateral remittance data, which is available only for a small set of countries, requiring extrapolation via calibration. Also, the disaster data from EMDAT, while comprehensive, excludes smaller-scale events that may nevertheless trigger localized remittance responses. Additionally, we do not consider income inequality among migrants \cite{antecol2000examination} and between migrants and locals \cite{amo2020migrant}, which can be relevant, while instead assuming that every remittance choice leads to sending a fixed share of the destination country's income per capita. This may oversimplify the diversity of remitting practices across diasporas. Lastly, our analysis focused solely on international remittances, but in several countries, internal remittances also play an essential role in migrants' livelihoods \cite{niimi2009determinants, housen2013systematic}. Notwithstanding these limitations, our analysis provides important and novel insights on bilateral remittance flows and the effect of disasters on these flows.
}

{
Future research could address the gaps associated with the limitations of our study by integrating household-level microdata, considering other sources of data on disaster events, and remittance flows. Regarding remittances data, developing consistent datasets on international and internal flows at high temporal granularity is crucial for uncovering new insights into the determinants of remittance decisions. Moreover, more work is needed in understanding the distributional effects of remittances, as their inflows to families of international migrants might increase inequality, thereby leaving the poorest households excluded \cite{koechlin2007international, anyanwu2011international}. Lastly, future research should also examine whether climate change intensification could alter the role of remittances in disaster resilience over the longer term.
}

{
Taken together, our findings confirm the importance of remittances as a private financial flow that responds systematically to disasters. Migrant diasporas act as informal insurers for their families and communities of origin, mobilising billions of dollars annually in response to disaster shocks. Yet, because these flows are concentrated in middle-income settings and taper off within months, they should be understood as complementary to, rather than substitutes for, collective mechanisms of disaster and climate finance.
}
}


\section{Methods}

\subsection{Model overview}

{
We develop a structural model that captures the dynamics of international migrants' remittance sending behaviour. We postulate that remittance sending is a binary choice made by every migrant $j$ at each period, with a time-dependent probability. Each international migrant is an agent in the model, and has a probability of sending remittances which evolves over time, determined by: i) their age-dependent ratio of earnings to consumption in the country of destination; ii) their probability of having a family in the country of destination; iii) the difference in GDP between origin and destination country; iv) the income level of the country of origin; v) the occurrence and magnitude of disasters in the origin country. If the outcome of the binary remittance choice is positive, the migrant remits a fixed portion of their monthly income. The model's parameters are calibrated to replicate a panel of monthly country-to-country remittance flows. The calibrated model is then used to extrapolate a global matrix of previously unobserved country-to-country remittance flows, including the effect of disasters (Figure \ref{fig:remittances_matrix}).

The building blocks for the model are three. First, the construction of a population of agents matching the population of international migrants globally. Second, the definition of a decision function, which determines if an when migrants send remittances to their countries of origin. Third, a reference dataset to calibrate the parameters in the decision function.
}

\subsection{Construction of the migrant population}
{
{
As a first modelling step, we build a synthetic population for all international migrants over the 2010-2019 period. A synthetic population is a simplified microscopic representation of the real target population, that is a collection of individual agents that has the same statistical characteristics of the real population \cite{chapuis2022generation}. For each pair of country of origin and destination of migrants, we aim to reproduce the age and gender distributions, meaning that we want to capture how many people from any country of a given sex and age cohort are living abroad in the period considered (see Supplementary Material, section \ref{sec:migrants}, for more details on the distribution of international migrants).
} 

{
To create the synthetic population of international migrants, we employ a proportional allocation method, referring to several data sources. The information on the aggregate stock of international migrants by gender, country of origin, and destination is taken from the United Nations (UN) International Migrants Stock \cite{UN_MigrantStock2024}. The UN dataset provides information on the stock of international migrants for the years 2010, 2015, and 2020. We interpolate the values to obtain monthly estimates using a cubic spline. To introduce an age dimension into these totals, we draw on more detailed census microdata available for selected countries with high-quality demographic reporting (United States \cite{USCensus_ACS2015}, Italy \cite{Ita_istat}, Germany \cite{destatis}, and Austria \cite{statcube}). From these censuses, we calculate the age distribution of migrants by sex, origin, and destination, and then average these distributions across the four reference countries. The resulting age profiles are subsequently applied to the aggregate sex-origin-destination totals in the broader dataset. This approach assumes that the age structures observed in the reference countries are broadly representative and can serve as proxies for contexts where age-specific migration data are unavailable.
}
}

\subsection{Remittance decision process}
{
{
To capture the evolution of each individual's probability of sending remittances, we operationalise the probability function using a logistic formulation. For an individual $j$ at time $t$ living in country $i$, we define the probability of sending remittances as
\begin{equation} \label{probabilityEqn}
    P_{j,i,t} = \frac{1}{1 + e^{-\theta_{j,i,t}}},
\end{equation}
where $\theta$, in turn, is a function of the variables that influence the remittance behaviour. In this model, we include five components: i) the agent's estimated ratio between income and consumption; ii) their probability of having family in the country of destination; iii) the wealth differential between the country of origin and of destination; iv) the income level of the country of origin, and v) the occurrence and magnitude of disasters in the country of origin. For individual $j$ at time $t$ living in country $i$, $\theta$ can be expressed as

\begin{equation}
\label{eq:theta}
\theta_{j,t} = 
\begin{cases}
\begin{aligned}
&\alpha + \beta_0 surplus_{j,i,t} + \beta_1 family_{j,i,t}  + \beta_2 \delta\_GDP_{j,i,t} + \beta_3 GDP\_capita_{j,t} \\
&\quad + \phi\, Disaster_{j,t},
\end{aligned}
& \text{if } surplus_{j,i,t} > 0 \\
-\infty, & \text{otherwise.}
\end{cases}
\end{equation}
}

\subsubsection{Earnings to consumption ratio}
{
The first variable we include, $surplus$, represents the migrant's income-to-consumption ratio and depends on age. This ratio is taken from the National Transfers Accounts (NTA), which compute the age profile for the domestic population's average earnings and consumption over a lifetime \cite{dalbis2015generational}. This variable is relevant because not all international migrants have access to the same financial resources, and age is a good proxy for financial resources. We expect the parameter $\beta_0$ to be positive, indicating that agents with a higher earnings-to-consumption ratio are more likely to remit. Agents without income, namely children below sixteen years of age, are assumed to have no possibility of sending remittances, or a probability of zero. Migrants' ages are determined during the construction of the synthetic population.}

\subsubsection{Family probability}
\label{sec:fam_prob}
{
The second factor we consider is the probability that an individual belongs to a household in the destination country. The intuition is that remittance motives differ for individuals who migrated alone compared to those who migrated with all or part of their household, with single individuals associated with a higher probability of sending remittances \cite{mahmud2020individual}. We operationalise this variable, $family$, by estimating the asymmetry of each origin-destination diaspora population's pyramid. More symmetric sex and age ratios in diasporas' population pyramids signal permanent settlement and family formation \cite{pedraza1991women, donato2011variations}. We define the asymmetry of a diaspora population based on its gender and age distribution in its population pyramid. For a given diaspora origin country $j$, destination country $i$, and period $t$, we define the age symmetry ($age$) as
\begin{equation}
  {age}_{j,i,t}
  = \frac{\min\!\left\{ parenting_{j,i,t},\, young_{j,i,t} \right\}}
         {\tfrac{1}{2}\left( parenting_{j,i,t} + young_{j,i,t} \right)},
\end{equation}
where $young_{j,i,t}$ is the number of people below 25 and $parenting_{j,i,t}$ is the number of people in parenting age, defined as being between 25 and 50 years old. We select equally wide age bins to define a balanced diaspora as one with equal numbers of young and parenting-age individuals. The age symmetry decreases if the number of young or parenting-age individuals outweighs the other within a given diaspora group.

Similarly, we define the sex symmetry ($sex$) as
\begin{equation}
  sex_{j,i,t}
  = \frac{\min\!\left\{ male_{j,i,t},\, female_{j,i,t} \right\}}
         {\tfrac{1}{2}\left( male_{j,i,t} + female_{j,i,t} \right)},
\end{equation}
where $male_{j, i,t}$ is the number of males in the diaspora population, and $female_{j, i,t}$ is the number of females. The sex symmetry shrinks for more skewed population pyramids along the sex axis. The pyramid asymmetry is defined by combining the sex and age symmetries according to
\begin{equation}
    {asymmetry}_{i,t} = 1- ({sex}_{j, i,t} \cdot {age}_{j, i,t}).
\end{equation}}

We take the asymmetry of the population pyramid along the sex and age axes to be a good proxy for the probability of migrants having a family in the destination country and establish that ${family}_{i,t} = {asymmetry}_{i,t}$. The rationale is that a skewed population pyramid, e.g. where everyone is overwhelmingly of one gender or one age category, indicates a higher likelihood of individuals having migrated alone. The family probability is calculated based on the demographic information on migrant diasporas.

\subsubsection{GDP and income}
{
The third variable we include in equation \ref{eq:theta}, $\delta\_ GDP$, is the differential in GDP per capita between the country of origin and the destination country. Remittance flows correlate with economic disparities: greater income gaps between host and origin countries drive higher remittance volumes \cite{yang2011migrant}. In this light, migrants moving from less wealthy countries have a stronger motive to send remittances to their country of origin \cite{carling2008interrogating, rapoport2006economics}. This assumption is also a key component of the estimation methodology for bilateral remittance flows used by the World Bank \cite{ratha2007south} (see Supplementary Material, section \ref{sec:KNOMAD}). We take GDP per capita data at the country level from the World Bank and normalise the relative difference. For migrant $j$ living in country $i$ at time $t$, define the relative difference as:

\begin{equation}
    {\delta\_ GDP}_{j,i,t} = 
    \begin{cases} (GDP_{i,t} - GDP_{j,t}) / GDP_{j,t} , & \text{if } GDP_{i,t} > GDP_{j,t} \\
    - (GDP_{j,t} - GDP_{i,t}) / GDP_{i,t},              & \text{otherwise},
    \end{cases}
\end{equation}

so that the ${\delta\_ GDP}_{j,i,t}$ takes values between -1 and 1, improving the tractability for the model. $GDP_{i,t}$ is the per-capita GDP of the country of destination, while $GDP_{j,t}$ is the per-capita GDP of the country of origin of the migrant.
} 

We also add the variable $GDP\_capita_{j,t}$ to equation \ref{eq:theta} to reflect the absolute income level per capita in the country of origin of migrant $j$. The logic is similar to the above, as migrants moving from rich countries have a lower probability of sending remittances. We operationalise this variable by applying min-max normalisation to each origin country's GDP per capita and we expect the variable to have a negative effect on the probability of sending remittances. 

\subsubsection{Disasters}
{For the component of the probability function \ref{eq:theta} which relates to disasters, we scale the magnitude of the disasters with a sinusoidal adjustment, such that

\begin{equation}
    \phi {Disaster}_{j, t} = \sum Disaster_{j,t} \cdot \Big( \text{height} + \text{shape} \cdot \sin\!\Big(\tfrac{\pi}{6} \cdot (t - x + \text{shift})\Big)\Big).
\end{equation}}
 
Therefore, we expect the occurrence of a disaster to affect remittance behaviour for several subsequent periods after the event. The magnitude of disasters $Disaster_{j,t}$ is formulated in terms of the percentage of a country's population that was affected by a given disaster. The number of people affected by disasters is taken from the Emergency Events Database (EMDAT, \cite{delforge2025dat}), while information on country populations is taken from the World Bank \cite{UNwpp24}. The parameters $height$, $shape$, and $shift$ control the intensity, oscillation, and temporal displacement of the disaster’s effect, allowing the model to capture both the immediate surge in remittance activity and the gradual dissipation of the effect in subsequent periods. The component $(t - x)$ indicates how many months have passed from the occurrence of the disaster. We estimate the effect of disaster on a window of 12 months. The oscillatory shape is in line with previous evidence on remittance behaviour, where it has been shown that an immediate surge in remittance sending can be compensated with a later reduction in remittance flows \cite{le2015poverty, bragg2018remittances}. The three parameters introduced here also allow for the possibility of disasters to show a null or negative effect in the short term, leaving the functional form agnostic to the impact we expect to see. The total impact of disasters for country of origin of $j$ at time $t$, $\phi {Disaster}_{j, t}$, is given by the joint impact of all disaster effects that overlap in a given month (see Supplementary Material, section \ref{sec:disaster_effect_SM}).
}

\subsubsection{Bilateral remittance flows}
{
To complete the model, we calculate remittance flows from country $i$ to country $J$ by aggregating the remittance decisions of individual migrants from country $J$ living in country $i$. To recap: each international migrant is modelled as an agent who, in each period, faces the decision to send or not remittances. This decision is stochastic and follows a Bernoulli process with probability $P_{j,i,t}$, which itself depends on i) their ratio of earnings to consumption; ii) their probability of having a family in the country of destination; iii) the difference in GDP between origin and destination country; iv) the income level of the country of origin; v) the occurrence and magnitude of disasters in the origin country. If the outcome of the Bernoulli choice is positive, the migrant remits a fixed fraction $\rho$ of their monthly income, approximated by the destination country's monthly per capita GDP, ${monthly\textunderscore GDP_{capita}}_{i,t}$. The estimated total flow is thus the expected value of the number of \enquote{successes} in this Bernoulli process, multiplied by the remittance amount per \enquote{successful} decision. The expected monthly flow of remittances from country $i$ to country $J$, called $R_{J,i,t}$, can be expressed as

\begin{equation}
    R_{J,i,t} \;=\; \mathbb{E}\!\left[ \sum_{n=1}^{N_{i,j}} Y_{n,t} \cdot \rho \cdot {monthly\textunderscore GDP_{capita}}_{i,t} \right],
\end{equation}
where $N_{i,j}$ is the total number of migrants from country $j$ residing in country $i$, and $Y_{n,t} \sim \text{Bernoulli}(P_{j,i,t})$ is the binary remittance decision of agent $n$ in period $t$, with success probability $P_{j,i,t}$. This formulation links individual-level remittance behavior to aggregate country-to-country flows, allowing calibration against observed panel data.
}

\subsection{Model calibration and validation}
\label{sec:rem_data}
{
The panel dataset used to calibrate the model combines monthly country-to-country remittance flow information from multiple sources. The dataset integrates both outflow and inflow records, depending on the data provider's reporting practices. Specifically, remittance outflows are available from Italy, capturing transfers from Italy to destination countries worldwide \cite{bettin2025responding}. In contrast, remittance inflows are reported by Mexico, Nicaragua, Guatemala, the Philippines, and Pakistan, reflecting transfers received from sending countries abroad. Where needed, the reported values have been converted to USD according to the exchange rate of the period. The complete panel dataset contains around 35 thousand monthly observations of bilateral remittance flows for a total of 769 billion USD over the 2010-2019 period. This is equivalent to around 12.5\% of the total remittance flows over the period. The dataset contains information on 194 unique sending countries and 140 unique receiving countries. The selection of these countries is constrained by the limited availability of monthly remittance data, as only a small number of national authorities publish remittance statistics at this frequency. The monthly dimension is particularly important for the analysis, as it enables the identification and estimation of time-sensitive impacts of disasters on remittance flows, which would be obscured in data aggregated at quarterly or annual intervals.
}

{
The model calibration is conducted by aligning simulated remittance flows with observed flows from the panel dataset of monthly country-to-country transfers. To avoid overfitting, the panel of observed remittances data is split into two samples, each containing 80 percent and 20 percent of the observations. Consequently, the aggregate monthly country-to-country remittance flows generated by the model for the training sample are compared to the empirical panel data, and the model parameters are adjusted to minimize the discrepancy between the two by minimizing the squared errors. The model parameters are calibrated using a gradient descent procedure that searches for the parameter set that minimizes the distance between observed and simulated remittances. The testing sample shows that more than 90\% of the variation in observed remittance flows can be captured by our model infrastructure (Figure \ref{fig:calibration_1}). This approach ensures that the model reproduces the key dynamics in the data and provides a reliable basis for extrapolating unobserved bilateral remittance flows across countries.

\begin{figure}[hbt!]
    \centering
    \includegraphics[width=\textwidth]{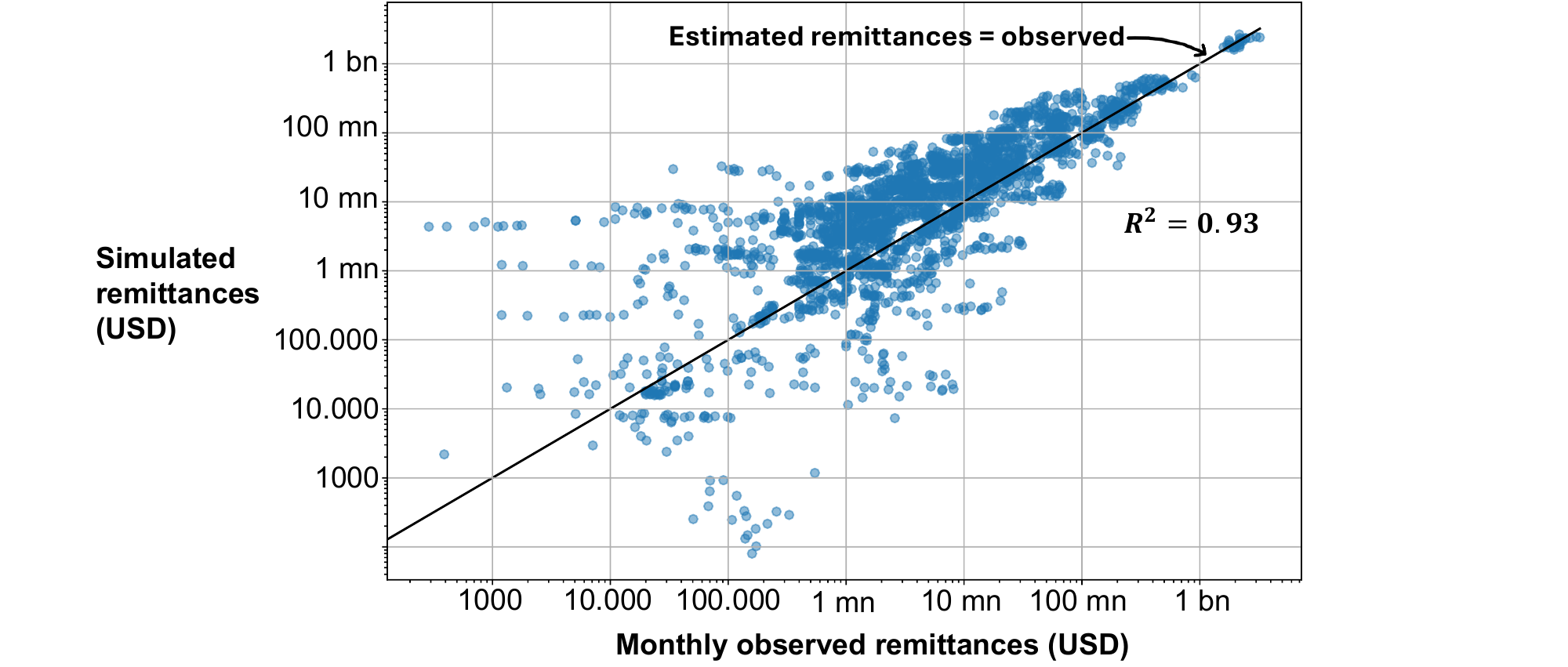}
    \caption{\textbf{Calibration results on the testing sample}. Comparison between the monthly bilateral remittances observed in the panel dataset and simulated by our model. The results are for observations that were not included in the calibration, representing a test sample.}
    \label{fig:calibration_1}
\end{figure}

}

\subsection{Estimating disaster-induced remittance flows}
{
To quantify the effect of disasters on remittances, we compute a counterfactual scenario using the calibrated model. Specifically, we re-run the exact parameterized model under a hypothetical environment in which no disasters occurred, while leaving all the other variables unchanged. The counterfactual remittance flows generated in this \enquote{no-disaster} scenario are used as a baseline to compare the estimates based on the real disaster data. The resulting differences isolate the portion of observed remittance dynamics attributable to disaster events within the calibrated model's bilateral flows estimates. The same procedure is repeated, isolating the occurrence of earthquakes, droughts, floods, and storms individually in order to isolate the impact of each disaster type. 
}


\newpage

\newpage
\appendix
\section*{Supplementary Material}

\renewcommand{\thesubsection}{\Alph{subsection}}

\subsection{Global distribution and demographics of international migrants}
\label{sec:migrants}
{
As of 2019, the global stock of international migrants reached approximately 272 million people, according to the United Nations’ International Migrant Stock \cite{UN_MigrantStock2024}. This corresponds to about 3.5\% of the world’s population at the time. The largest group consists of migrants from upper-middle-income countries (35\% of the total), while the smallest group is migrants from low-income countries (12\% of the total). The distribution of international migrants by country of destination is also not homogeneous: 72\% of international migrants live in high-income countries, where they make around 14\% of the total population \cite{un_migration_2019}. Only 3.8\% of international migrants live in low-income countries. Indeed, most international migrants have moved to richer countries, with the two largest groups being migrants from lower- and upper-middle-income countries now living in high-income countries (Table \ref{tab:migration}). The third largest group remains migrants who moved between high-income countries. Migrants from low-income countries tend to be evenly distributed across destination-country income groups. 

Breaking down international migration by gender shows that women comprised just under half of the global migrant population in 2019. This share is in slight decline from around 49\% in 2000 to 47.5\% in 2019. Also, for the gender breakdown, the shares are not homogeneous across migrant groups. In fact, around 51\% of the international migrants from high and upper-middle-income countries are female. For migrants from lower-middle and low-income countries, the shares are much lower (44\% and 41\% respectively). 

In terms of age, international migrants are predominantly of working age. In 2019, 74\% of all international migrants were between the ages of 20 and 64, compared to 57\% of the global total population \cite{un_migration_2019}. Conversely, only one in seven international migrants was under the age of 20, compared with one in every four persons globally \cite{un_migration_2019}. International migrants living in high-income countries tend to be older. In 2019, the median age of migrants in Northern America (43.5 years) and Europe (42.7 years) was higher than the global median (39 years). By contrast, international migrants coming from and living in lower-income countries tend to be younger, with a median age significantly lower than the global average. As discussed in section \ref{sec:fam_prob}, the demographic composition of migrant diasporas is one of the determinants of remittance behaviour.
}

\begin{table}[htbp!]
\centering
\begin{tabularx}{\textwidth}{l *{8}{>{\centering\arraybackslash}X}}
\hline
\multicolumn{1}{l}{Destination} & \multicolumn{2}{c}{High income} & \multicolumn{2}{c}{\makecell{Upper middle\\income}} & \multicolumn{2}{c}{\makecell{Lower middle\\income}} & \multicolumn{2}{c}{Low income} \\ 
\cline{2-9}
\multicolumn{1}{l}{Origin} & male & female & male & female & male & female & male & female \\ 
\cline{2-9} 
High income & 20,77 & 21,68 & 3,92 & 4,65 & 0,97 & 1,13 & 0,04 & 0,04 \\
Upper middle income & 45,59 & 47,95 & 4,18 & 3,86 & 1,34 & 1,51 & 0,16 & 0,15 \\
Lower middle income & 35,33 & 23,12 & 6,94 & 5,63 & 7,15 & 6,29 & 1,49 & 1,53 \\
Low income & 5,14 & 2,90 & 4,63 & 3,54 & 5,54 & 4,88 & 3,71 & 3,78 \\ 
\hline
\end{tabularx}
\caption{Stock of international migrants in 2019 by income group of origin, income group of destination, and sex, in millions of people.}
\label{tab:migration}
\end{table}

\subsection{Disaster exposure and impacts}
\label{sec:disasters}
{
The decade between 2010 and 2019 was marked by over 3,000 weather- and climate-related disasters. Floods and storms were the most frequent events and together accounted for the bulk of recorded disasters, while droughts and earthquakes produced some of the largest slow-onset and sudden losses, respectively \cite{cred2020human}. The major single-event catastrophes in the decade by death toll were all sudden-onset disasters and included the 2010 Haiti earthquake, the 2011 Tōhoku earthquake–tsunami in Japan, and the 2015 Nepal earthquake. Large storm and flood events such as the 2013 Typhoon Haiyan also characterised the decade and produced extremely high local death tolls and displacement.

 \begin{table}[h!]

\begin{tabularx}{\textwidth}{l| *{8}{>{\centering\arraybackslash}X}}
\multicolumn{1}{l}{}                       & \multicolumn{4}{c}{People affected (\% of which dead)}                                                               \\
\multicolumn{1}{c|}{\textbf{}}             & \multicolumn{1}{c}{Drought} & \multicolumn{1}{c}{Earthquake} & \multicolumn{1}{c}{Flood} & \multicolumn{1}{c}{Storm} \\ \hline
\multicolumn{1}{c|}{High income}           & 0.9 (0.05\%)                & 5.8 (0.37\%)                   & 7.7 (0.04\%)              & 93.7 (0.01\%)             \\
\multicolumn{1}{c|}{Upper middle   income} & 196.0 (\textless{}0.01\%)   & 28.5 (0.23\%)                  & 502.1 (\textless{}0.01\%) & 103.9 (0.02\%)            \\
\multicolumn{1}{c|}{Lower middle   income} & 459.3 (\textless{}0.01\%)   & 18.9 (1.27\%)                  & 310.0 (0.01\%)            & 215.5 (0.01\%)            \\
\multicolumn{1}{c|}{Low income}            & 261.7 (0.01\%)              & 9.8 (0.10\%)                   & 35.7 (0.04\%)             & 17.9 (0.02\%)            
\end{tabularx}
\caption{Total number of people affected (in millions) and dead (in percentage) by disasters between 2010 and 2019, by income group of the affected country and disaster type.}
\label{tab:disasters}
\end{table}

Droughts during the 2010–2019 period also affected a very large number of people, far more than many sudden-onset disasters, but tended to produce a relatively lower death toll (Table \ref{tab:disasters}). Over the 2010 to 2019 period, more than 900 million people were estimated to have been impacted by drought. By contrast, the WMO’s Atlas of Mortality and Economic Losses (1970–2019) records only about 650,000 deaths due to droughts over that entire half-century \cite{douris2021atlas}. This mismatch, high exposure but fewer deaths, reflects the nature of drought as a \textit{creeping phenomenon}: it tends to erode livelihoods and food security over time rather than trigger immediate mass-casualty events. Earthquakes show a higher death rate than all other disasters, especially for events that took place in lower-middle-income countries.

 Impacts were markedly unequal by country income group, particularly in terms of exposure. Low- and lower-middle-income countries account for the largest numbers of people affected by floods, storms, and droughts, reflecting higher vulnerability and population exposure. Differences in reported mortality rates across income groups are comparatively small and do not follow a clear pattern, especially for floods and storms. However, mortality and economic losses are likely underreported in lower-income countries, and death rates alone do not capture the heavier disruption to health systems, livelihoods, and the slower recovery observed in these contexts. By contrast, high-income countries tend to register larger insured and recorded economic losses but relatively low mortality, reflecting stronger infrastructure and preparedness \cite{nashwan2023impact}.

 Finally, while floods, storms, earthquakes, and droughts accounted for most events and much of the damage in 2010–2019, other hazards, which we did not include in our analysis, also produced large death tolls and escalating economic costs - notably heatwaves. During the 2010s, heatwaves emerged as one of the deadliest climate extremes, with average global excess deaths from heat estimated at around 153,000 per warm season \cite{zhao2024global}. This rise was driven by both increasing ambient temperatures and more frequent prolonged high-temperature spells, as climate change made oppressive heat more common. In 2019 alone, record-breaking heat waves hit regions from Europe to India, Japan, and Australia. For example, a late-July heat surge in central and western Europe was linked to nearly 3,000 excess deaths in the Netherlands. Vulnerable populations such as the elderly and people with pre-existing health conditions were disproportionately affected, since heat stress exacerbates cardiovascular, respiratory, and other illnesses.

 Heatwaves were not included in the analysis because estimation of their impacts differ fundamentally from those of floods, storms, earthquakes, and droughts. Heatwaves' impacts are primarily measured through excess mortality and health-system records rather than a clearly defined affected population or economic damages. Because excess-death estimates vary widely by methodology and are often produced retrospectively, they introduce inconsistencies that would reduce comparability across hazards. In contrast, floods, storms, earthquakes, and droughts have more standardized reporting frameworks and more reliable event-level data, allowing for a coherent cross-hazard comparison.
}

\subsection{World Bank (KNOMAD) remittance estimates}
\label{sec:KNOMAD}
{

The main provider of data on international remittance flows is the Global Knowledge Partnership on Migration and Development (KNOMAD). KNOMAD is a World Bank-led initiative that serves as a global hub for knowledge on migration and development issues. Its mission is to produce and synthesize knowledge and to offer technical assistance to improve migration policies and their impact on development, with the production of yearly reports on remittance flows being one of its key products.

KNOMAD produces two datasets on remittances, both containing observations at the yearly level. The first collects information on total remittances inflows and outflows at the country level, while the second is a matrix of bilateral remittance flows between countries \cite{KNOMAD_Remittances}. Both datasets rely on different estimation methodologies. 

The total remittance inflows, representing the aggregate amount of money a country receives from and sends to all other countries combined, are typically compiled based on the country's Balance of Payments (BoP) statistics, primarily drawing from data reported by central banks and based on international guidelines (like the IMF's BoP Manual). This estimation suffers from structural issues in reporting by the competent authorities, leading to reported inflows totaling more than 1.5 times the reported outflows.

The bilateral remittance data are estimated using a gravity model framework \cite{ratha2007south}. This approach uses available data on the stock of international migrants and assumes the average remittance size sent by a migrant is determined by the per capita GDP of the origin and destination countries. According to this methodology, the average remittances sent by a migrant from country $j$ living in country $i$ is defined as:

\begin{equation}
    r_{i, j} = f(Y_j, Y_i) = \begin{cases}
  Y_j, & \text{if } Y_i < Y_j, \\
  Y_j + (Y_i - Y_j)^\beta, & \text{if } Y_i \geq Y_j
\end{cases},
\end{equation}
where $Y_i$ is the per capita GDP of the destination country and $Y_j$ is the average per capita GDP of the country of origin. Consequently, the total remittance flow to country $j$ is 
\begin{equation}
    R_{j} = \sum_i r_{i,j} M_{i,j},
\end{equation}
where $M_{i,j}$ is the total number of migrants from country $j$ living in country $i$. 

\subsubsection{Comparison of results with World Bank estimates}

To strengthen the relevance of our model, we conduct an additional test by comparing its performance with that of the KNOMAD model for estimating bilateral remittance flows. We do so by comparing the estimates from both models with the panel of observed monthly bilateral remittance flows for which we collected primary data (Table \ref{tab:remittance_compare}). We calibrate the exponent $\beta$ in the KNOMAD model to minimise the sum of squared errors relative to the data.  

We verify that, on average, the relative error per bilateral flow of our model is 45\% of the KNOMAD model's average relative error. In particular, the KNOMAD methodology largely overestimates remittance flows between the USA and Mexico, as well as several flows between high-income countries in Europe, such as between Italy and Romania. On the contrary, our model overestimates the flows from the United Arab Emirates and Saudi Arabia to Pakistan, as well as the flows from Canada to the Philippines.

A core component of the performance difference is due to the flexibility of our model, which is able to include other diaspora characteristics beyond the raw number of individuals in a diaspora group and the income difference between the two countries. On the contrary, the model employed by KNOMAD cannot differentiate between a diaspora composed solely of children and one of solely working-age adults. Overall, this issue leads the KNOMAD methodology to often overestimate remittance flows (Figure \ref{fig:calibration_3}). The comparison with the KNOMAD estimates shows that the structural model we developed can be a credible alternative to the model employed by the World Bank for estimating bilateral remittance flows.

\begin{figure}[h!]
    \centering
    \includegraphics[width=\textwidth]{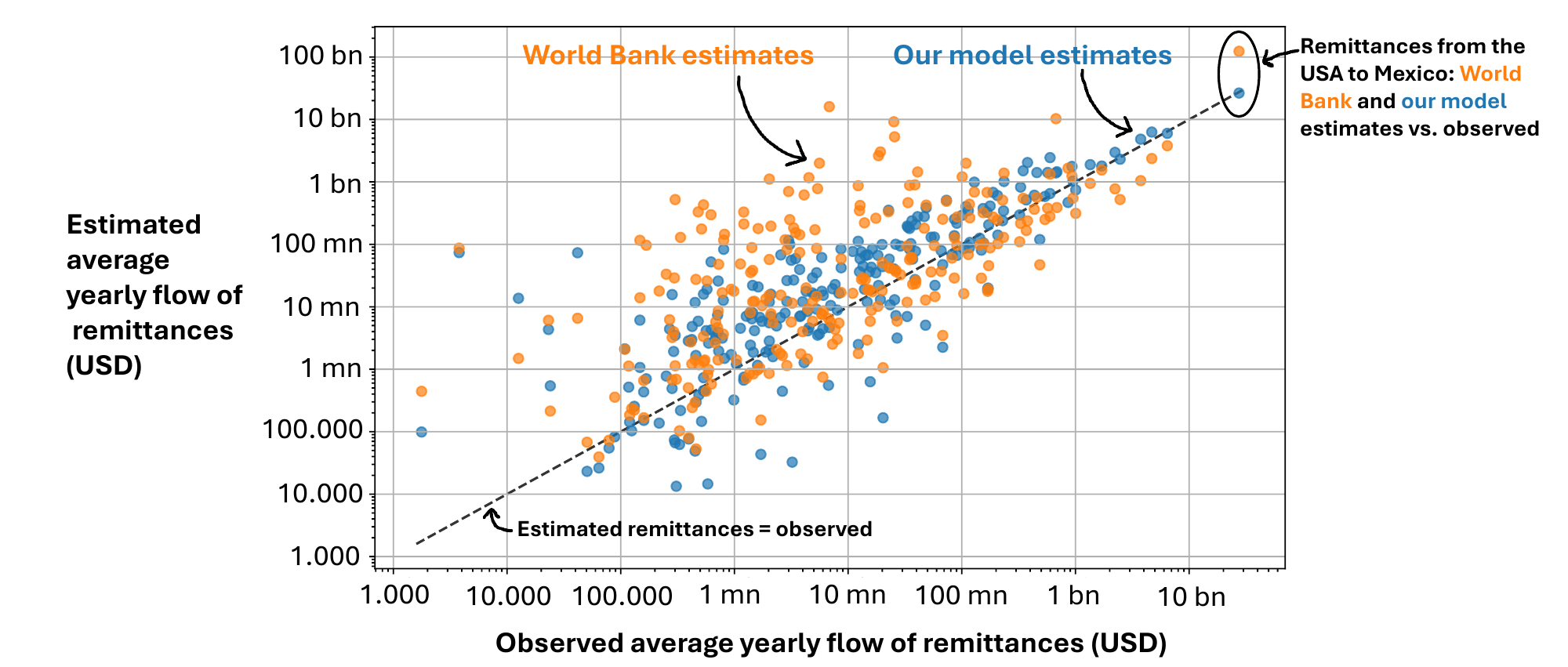}
    \caption{\textbf{Comparison of estimated bilateral flows with World Bank estimates.} Yearly average observed remittance flows between a pair of countries (x-axis), and estimates of the same flows obtained using our model (in blue, y-axis) and the World Bank's model (in orange, y-axis). All values are in billions of USD.}
    \label{fig:calibration_3}
\end{figure}

\newpage

\begin{table}[h!]
\centering
\begin{tabularx}{\textwidth}{
    l l 
    >{\centering\arraybackslash}X 
    >{\centering\arraybackslash}X 
    >{\centering\arraybackslash}X 
    >{\centering\arraybackslash}X 
    >{\centering\arraybackslash}X
}
\textbf{Sender} &
\textbf{Recipient} &
\textbf{Observed} &
\textbf{Our model estimate} &
\textbf{WB model estimate} &
\textbf{Our model square error} &
\textbf{WB model square error} \\
\hline
USA & Mexico & 27,46 & 26,49 & 123,28 & 0,95 & 9181,54 \\
USA & Guatemala & 6,42 & 5,98 & 3,80 & 0,19 & 6,90 \\
Saudi Arabia & Pakistan & 4,69 & 6,28 & 2,37 & 2,54 & 5,39 \\
UAE & Pakistan & 3,75 & 4,83 & 1,05 & 1,18 & 7,29 \\
USA & Pakistan & 2,47 & 2,31 & 0,52 & 0,03 & 3,79 \\
UK & Pakistan & 2,23 & 2,97 & 0,77 & 0,56 & 2,12 \\
Saudi Arabia & Philippines & 1,70 & 1,79 & 1,55 & 0,01 & 0,02 \\
UAE & Philippines & 1,35 & 1,89 & 0,95 & 0,29 & 0,16 \\
United Kingdom & Philippines & 1,01 & 0,74 & 0,31 & 0,07 & 0,48 \\
Japan & Philippines & 0,95 & 1,03 & 0,55 & 0,01 & 0,16 \\
Italy & Philippines & 0,93 & 1,77 & 1,24 & 0,70 & 0,10 \\
Italy & China & 0,86 & 0,47 & 1,66 & 0,15 & 0,63 \\
Kuwait & Pakistan & 0,69 & 1,45 & 0,39 & 0,59 & 0,09 \\
Italy & Romania & 0,67 & 1,39 & 10,28 & 0,52 & 92,25 \\
Canada & Philippines & 0,60 & 2,45 & 1,33 & 3,42 & 0,54 \\
\end{tabularx}
\caption{Comparison between average yearly observed remittances, model estimates, and World Bank estimates, with squared errors. Values are in billion USD. The 15 largest remittance channels from the panel dataset are displayed.}
\label{tab:remittance_compare}
\end{table}
}

\newpage
{
\subsection{Calibrated model parameters}
We report here the parameters resulting from the calibration of our model (Table \ref{tab:params}). The direction of the effect for each calibrated parameter corresponds to the hypothesised direction described in the model. A higher earnings-to-consumption ratio and a larger difference in GDP per capita between the origin and destination countries increase the likelihood of remitting. Conversely, a higher probability of having family in the country of destination and higher income levels in the country of origin decrease the probability of remitting. Regarding the impact of disasters, we estimate that a disaster immediately increases the remittance probability, with the largest increase occurring between three and four months after the disaster. This is in line with previous evidence \cite{bettin2025responding}. The effect peters out after nine months, with a small compensatory effect afterwards. Lastly, an average migrant is estimated to send around 18 percent of the host country's monthly average income per capita when they remit.

\begin{table}[h!]
\centering
\begin{tabularx}{\textwidth}{l c c c X}
\textbf{Parameter} &
\textbf{Value} &
\textbf{95\% CI Lower} &
\textbf{95\% CI Upper} &
\textbf{Description} \\
\hline

$\alpha$        & 0.02  & -0.02 & 0.03 & Constant term in remittance probability function \\

$\beta_0$       & 1.08  & 1.03  & 1.20 & Effect of earnings-to-consumption ratio \\

$\beta_1$       & -4.65 & -4.76 & -4.56 & Effect of family probability in destination country \\

$\beta_3$       & 2.83  & 2.69  & 2.95 & Effect of GDP difference between origin and destination \\

$\beta_4$       & -3.67 & -3.56 & -3.74 & Effect of GDP per capita of the country of origin \\

\text{height}   & 0.15  & 0.05  & 0.19 & Baseline level of disaster impact function \\

\text{shape}    & 0.19  & 0.11  & 0.24 & Amplitude of disaster impact oscillations \\

\text{shift}    & -0.98 & -0.99 & -0.72 & Temporal displacement of disaster impact function \\

$\rho$          & 0.18  & 0.17  & 0.19 & Fixed portion of monthly income remitted \\
\hline
\end{tabularx}
\caption{Table of estimated model parameters with 95\% confidence intervals.}
\label{tab:params}
\end{table}

}

\subsubsection{Nonlinearities in disaster-induced remittance responses}
\label{sec:disaster_effect_SM}
{
A key feature of the structural model developed in this paper is that individual migrants' response in remittance behaviour depends both on the magnitude of the disasters as well as the pre-disaster remittance probability for the migrant. Every migrant in the model has a $\theta$ score, which determines their probability of sending remittances (Equation \ref{eq:theta}). The disaster impact has the same effect on the $\theta$ score of each migrant irrespective of the other factors which influence remittance behaviour (Figure \ref{fig:disasters_estimate}, A).

The increase in $\theta$ score translates into heterogeneous changes in the effective remittance probability for each migrant, depending on their pre-disaster probability. A migrant with high pre-disaster probability of sending remittances will only slightly increase their probability of sending money after the disaster (Figure \ref{fig:disasters_estimate}, B). Similarly, a migrant with very low pre-disaster probability of sending remittance will only slightly change their behaviour. On the contrary, individuals with median probability of sending remittances show the highest capacity to activate their remittance sending in response to a disaster. This non-linearity in the effect of disaster is a conceptual advancement of our model which allows to capture heterogeneities in diaspora remittance responses to disasters.

\begin{figure}[h!]
    \centering
    \includegraphics[width=\textwidth]{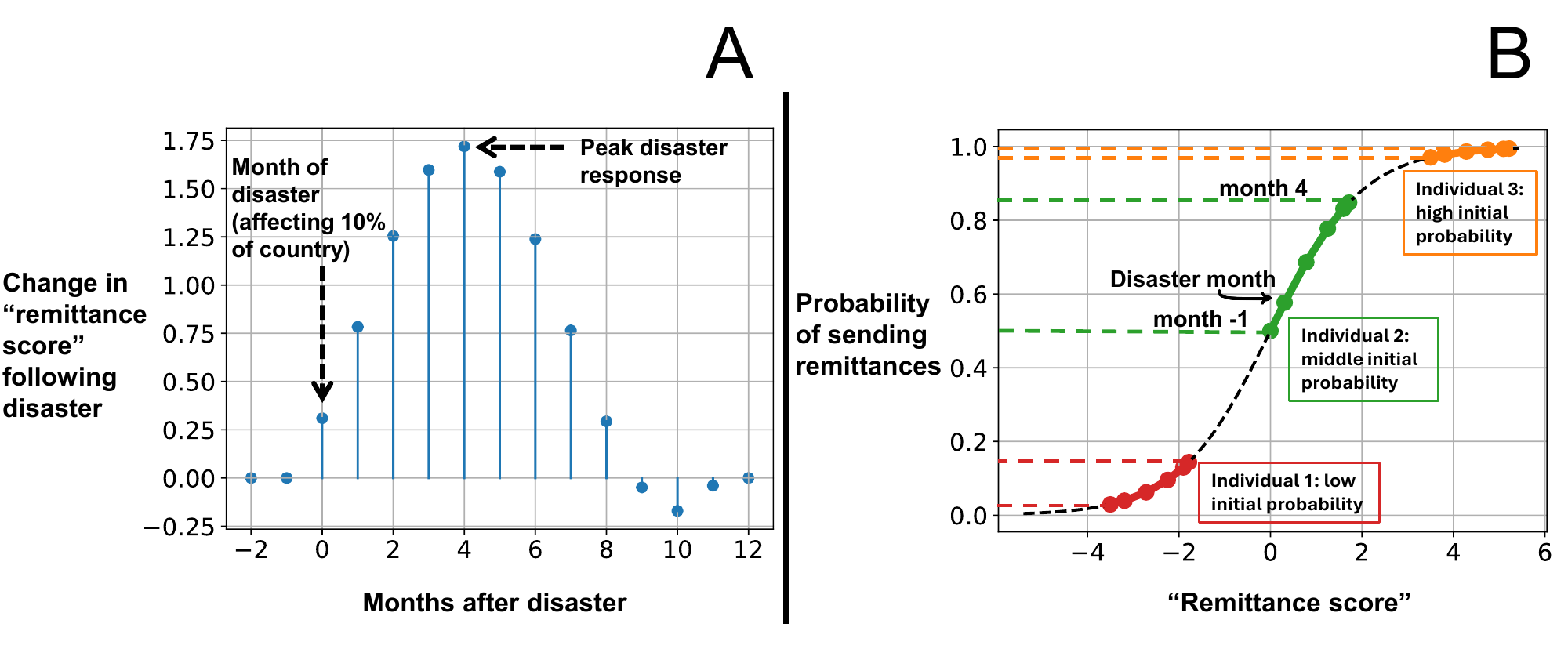}
    \caption{\textbf{Effect of disasters on individual remittance probabilities.} Panel A: effect of a disaster affecting 10\% of a country's population on the \enquote{remittance score} $\theta$ (y-axis) of a migrant from the same country over 12 months (x-axis). Panel B: change in migrant's probability of sending remittances in the four months after a disaster, depending on their pre-disaster probability of sending remittances.}
    \label{fig:disasters_estimate}
\end{figure}
}

\subsubsection{Uncertainty due to stochastic remittance decisions}


The structural model we developed relies on a stochastic process for remittances decisions, where every international migrant makes a Bernoulli choice at every period about whether to send money or not. However, the calibration strategy for the model relies on the deterministic formulation of the problem, where the total number of remittance senders is given by the sum of the expected values of all the Bernoulli choices. To quantify the uncertainty connected to the stochastic nature of the model, we construct 95\% confidence intervals for the simulated remittance flows. Each international migrant in our model is associated with a probability of remitting in a given month, and makes a Bernoulli decision at each period based on this probability. The aggregate remittance activity is obtained by summing these independent Bernoulli trials. This process produces a Poisson-binomial distribution for the total number of senders, with mean equal to the sum of individual probabilities and variance determined by the sum of the corresponding $P_{i,j,t}(1-P_{i,j,t})$ terms. 

\begin{figure}[hbt!]
    \centering
    \includegraphics[width=\textwidth]{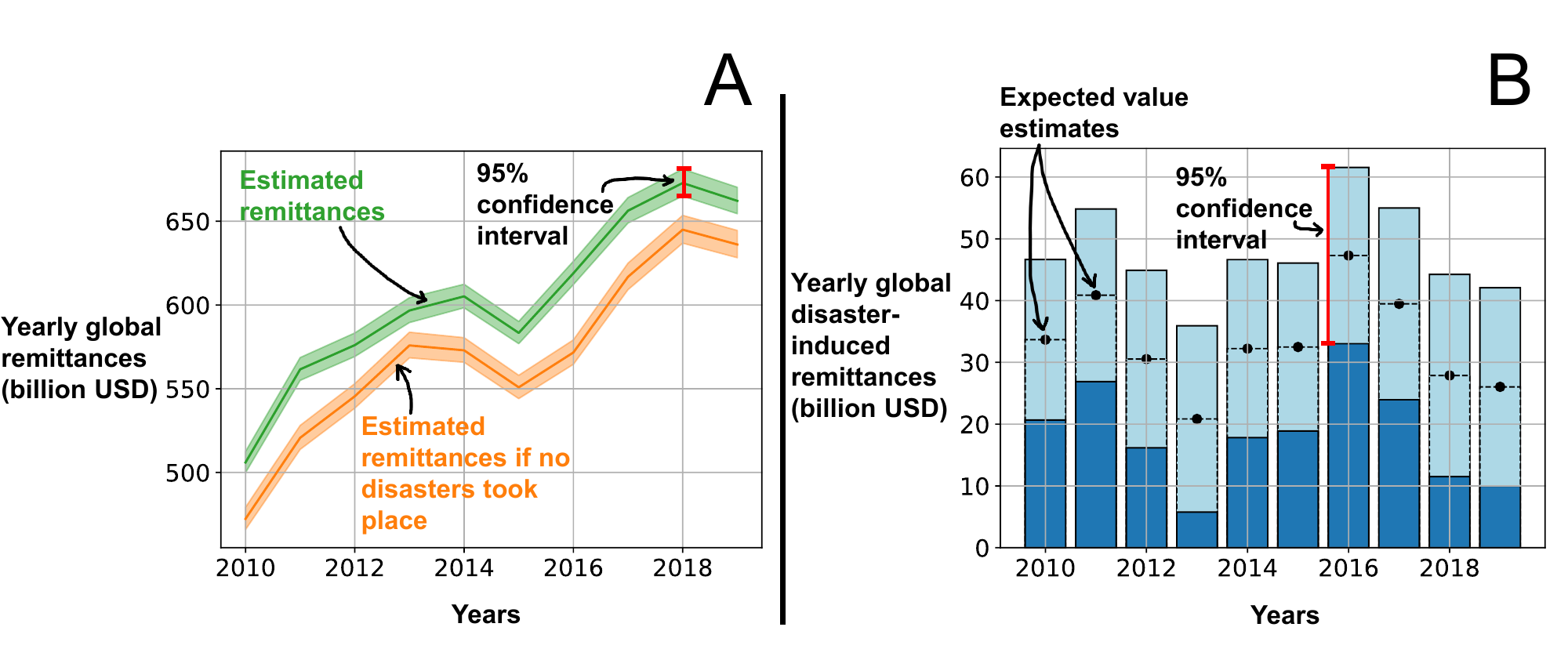}
    \caption{\textbf{Confidence interval for estimated remittance flows and disaster-induced remittances.} Panel A shows the 95\% confidence interval over time for the estimated remittances using our model (y-axis, billion USD), as well as for the counterfactual scenario where no disasters took place. Panel B reports the 95\% confidence interval for the yearly total disaster-induced remittances  (y-axis, billion USD). The blue columns represent the most-conservative estimate for the yearly disaster-induced remittances.}
    \label{fig:uncertainty}
\end{figure}

We propagate this uncertainty into the aggregate remittance flows by repeatedly sampling from the implied Poisson-binomial distribution and computing the resulting distribution of simulated remittances (Figure \ref{fig:uncertainty}, A). The 95\% confidence intervals reported correspond to the 2.5th and 97.5th percentiles of these simulations and represent upper and lower confidence bounds for total remittance flows over time. This approach faithfully reflects the heterogeneity in individual-level remittance probabilities and provides a transparent measure of the model’s estimation uncertainty. We repeat the procedure for the counterfactual scenario where no disaster took place in order to obtain confidence intervals for the total disaster-induced remittances (Figure \ref{fig:uncertainty}, B).

\end{document}